# Estimation of the surface-d to bulk-s Crossover in the Macroscopic Superconducting Wavefunction in Cuprates.

K.A. Müller, Physics Institute, University of Zürich, Winterthurerstr. 190, CH-8057 Zürich, Switzerland.

Abstract: The concept of a surface d- and bulk s-symmetry in cuprate superconductors is applied to recent small-angle neutron-scattering results. These show a change of hexagonal to square vortex lattice as a function of the magnetic field along the c-axis. Identifying the hexagonal lattice with s- and the square with d-symmetry, the crossover distance from the surface d to the bulk s perpendicular to the c-axis is estimated to be 35 nm for LSCO and roughly 7 nm for YBCO, both at optimum doping. The crossover along the c-axis has to be of only a few layers distance to reconcile tunneling, photoemission and pulsed femtosecond reflectivity experiments. These estimates are compatible with µ-rotation, NMR and other experiments.

A large part of the community considers the macroscopic superconducting wavefunction in the cuprates to be of d-symmetry. Pertinent evidence has been obtained by experiments in which mainly surface phenomena have been used, such as tunneling or the well known tricrystal or tetracrystal experiments [1]. However, recently, data obtained by probing the property in the bulk has yielded increasing evidence that inside the cuprate superconductor a substantial s-component is present, and therefore I have proposed a changing symmetry from purely d at the surface to more s inside, at least [2]. This suggestion was made to reconcile the observations stemming from the surface and bulk. But such a behavior would be at variance with the accepted classical symmetry properties in condensed matter [1,3]. In this respect, Iachello, applying the interacting boson-model successful in nuclear theory, to the $C_{4v}$ symmetry of the cuprates, showed that indeed a crossover from a d-phase at the surface, via a d +s, to a pure s-phase should be present [4]. The question thus arose at which distance from the surface this crossover occurs. Using recently published small-angle neutron-scattering (SANS) data in LSCO, in which an intrinsic square vortex-lattice exists at fields larger than 0.4 Tesla [5], we estimate here the crossover distance in question to be $R^c_{ab}$ = 350 Å for this material. In YBCO, on the other hand, unpublished SANS data [6] indicate that $R^c_{ab}$ is five times smaller than in LSCO. We then show that this distance is compatible with earlier SANS and decoration data in the first compound. Furthermore, the d-symmetry deduced from nuclear magnetic resonance (NMR) also fits with our analysis, and µ-rotation experiments are compatible as well. The crossover distance along the c-axis, $R^c_c$, appears to be much shorter than $R^c_{ab}$, only one or two unit-cell distances, considering pulsed femtosecond experiments in YBCO and photoemission studies in BSCCO.

A clear observation of a square vortex lattice with a magnetic field along the c-axis in a single crystal of YBCO is now ten years old [7]. Most recently the group of Mesot [5], in a remarkable SANS experiment, observed a crossover from a hexagonal to a square vortex lattice in a single crystal of $La_{1.83}\,Sr_{0.17}\,CuO_{4+\delta}$. This occurred upon enhancing the external magnetic field directed along the c-axis of the crystal used. In Fig.1 the ratio of hexagonal to square intensity of the SANS diffraction spots as a function of the magnetic field is reproduced (from Fig. 3b of Ref. 5 and Fig 4. of Ref. 8). Whereas at low field, the hexagonal intensity is predominant, it is reduced and levels off near $B_c = 0.4$ Tesla. From there on the square lattice prevails. Such a lattice has been predicted theoretically to occur for d-wave symmetry of an superconductor, with the squares oriented along Cu-O-Cu bonds, i.e. spots along {1,1} or {1,-1} [9,10], as observed, or rotated by $45^0$ around the c-axis [11,12]. From these theories and the SANS experiment, we can assume that above 0.4 Tesla the symmetry of the macroscopic wavefunction is predominantly d. Below this field, the wavefunction crosses over to s-symmetry because for this symmetry the vortex lattice is hexagonal, as is well known [13].

From the above consideration we can directly calculate the crossover distance in the ab plane, $R^{ab}_c$, of the postulated [2] surface d- to bulk s-symmetry: In the SANS experiment at low magnetic fields, the vortex distance is larger than $2R^{ab}_c$. Thus the bulk symmetry has to be s from the hexagonal spots. At high fields, the vortex distance is smaller than $2R^{ab}_c$, and the area occupied by the superfluid is almost all d. The factor of 2 occurs because the distance in question is between two normal conducting cores, i.e. twice the distance between the two cores. The critical magnetic field, where it is to 80% d-wave, is $B_c = 0.4$ Tesla as obtained from Fig. 1. Thus we can estimate $R^{ab}_c$: The flux through the ab-plane is $B_c = n\,\phi_0$, where n is the number of vortices per unit area, and $\phi_0 = h/2e = 2\,10^{-7}\,Gauss.cm^2$ is the flux quantum. For a square vortex lattice, n is simply $(2R^{ab}_c)^{-2}$. Thus for $R^{ab}_c$ we get:

$$R^{ab}_c = \tfrac{1}{2}(\phi_0/B_c)^{1/2} = 0.35\,10^{-5}\,cm = 35\,nm\,or\,350\,\text{Å}.$$

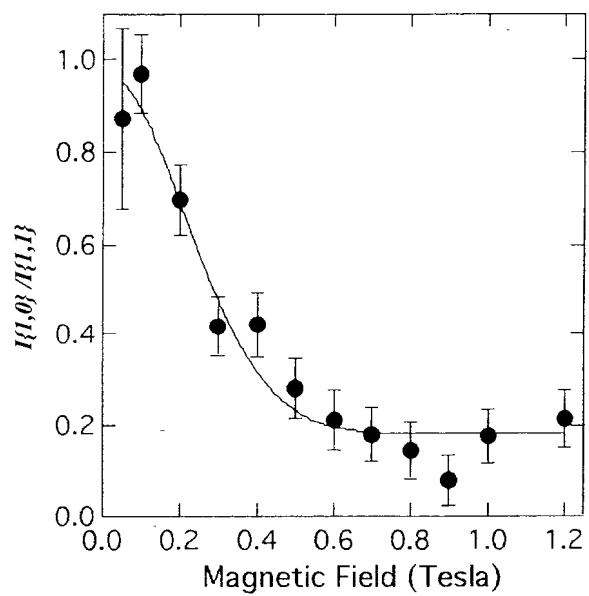

FIG.1: Ratio of the hexagonal {1,0} to quadratic {1,1} intensity spots of SANS in a single crystal of $La_{1.83}\,Sr_{0.17}\,CuO_{4+\delta}$ as a function of a magnetic field along the c-axis. The leveling off near 0.4 Tesla is marked. Adapted from Fig.4 of Ref. 8 and Fig. 3b of Ref. 5

In YBCO most recent SANS experiments indicate that the crossover occurs at a much higher field namely between 8 to 9 Tesla [6]. If this observation is correct then the above analysis puts $R^{ab}_c$ at 70 Å. The early observation of a square vortex lattice in YBCO [7] was attributed to flux pinning at the time. The applied magnetic field was 0.8 Tesla, ten times smaller than the $B_c$ for YBCO. Thus, this conclusion appears to have been correct because for $B << B_c$ a hexagonal lattice has to be present in the unperturbed crystal. This was the case in decoration experiments in YBCO, which showed a hexagonal lattice in a low magnetic field with flux distance of ~100 nm [14]. This distance 140 times larger than the above estimate of $R^{ab}_c$ for YBCO. The hexagonal lattice observed has to be attributed, in our view, entirely to bulk s-wave superconductivity.

With the above notion, we can now turn our attention to the NMR. The data have been analyzed by Scalapino, assuming a homogenous superfluid [15]. This analysis is probably correct owing to the long time window of the NMR experiments of $10^{-6}$ to $10^{-7}$ sec. However, at short times, the $CuO_2$ plane is heterogeneous in spin, charge and lattice displacements [16]. Scalapinos analysis was largely in favor of d-wave symmetry, but the experiments were all conducted with modern NMR spectrometers employing magnetic fields of 6 to 9 Tesla. As noted above this is in the range of $B_c$, where the flux lattice is so dense that it induces at a large overall d-symmetry. Much effort has been devoted to µ-rotation experiments to decide on the superfluid symmetry present in the copper oxides. Basically in this type of experiments one probes the muon- spin depolarisation rate, which is inversely proportional to the square of the London-penetration depth $\lambda_r(T)$ [17]. This provides information on the decay of the magnetic field inside the superconductor. In YBCO at low temperatures, $\lambda_{ab}(o) = 1500$ (200) Å at near optimum doping [17]. Thus $\lambda_{ab}(T)$ probes a near bulk property; moreover $\lambda_{ab}(T)$ increases with temperature. For a fully developed gap in all directions, one expects a vanishing slope for $\lambda_{ab}(T) \rightarrow \lambda_{ab}(0)$, which was indeed observed in the first careful experiments [18,19] using YBCO powders. In later experiments carried out with single crystals of good quality, a linear slope of $\lambda_{ab}(T) \rightarrow \lambda_{ab}(0)$ was reported, as expected for d-symmetry [20,21]. However most recently Harshman et al. [23] showed, with a very careful analysis of the depolarisation rate as a function of temperature and various magnetic fields, that the observed linear term in $\lambda_{ab}(T)$, results from flux depinning in the high-quality single crystals. In the early experiments [18,19] the vortices were apparently pinned because of the much poorer quality as compared with those of single crystals available now. It should be noted that all µ-rotation experiments discussed in this paragraph, and in contrast to NMR, were carried out in magnetic fields B *well below* the critical magnetic field $B_c$, where the superfluid symmetry changes into d character.

So far we have considered the crossover distance $R_{ab}^c$ along an a- or b-direction in the crystal, i.e. perpendicular to the c-axis. Owing to the layered structure of the cuprates it is highly probable that the crossover along the c-direction $R_c^c$ will be different. At present we do not have experiments from which we can deduce $R_c^c$ as was done above for $R_{ab}^c$. But there exist new data from which we can estimate it: The optical penetration depth in YBCO along the c-axis is approximately 1500 Å [24]. Therefore we expect that such optical experiments will "see" beyond $R_c^c$ inside the crystal, and detect an s-symmetry. This is indeed the case. For a fully developed gap, as for s-symmetry, the change in reflectivity, ΔR/R, after a pair-breaking first pulse, and measured by a probing second pulse, should be proportional to the fluency F of the first pulse. This is so because the number of broken Cooper pairs, responsible for the reflectivity change, is then proportional to F [24]. The reflectivity data of optimally doped YBCO obtained by two optical groups [25,26] show a strict proportionality of ΔR/R and F over two orders of magnitude (see Fig. 2). Were a global d-gap present, one would expect ΔR/R ∝ F$^{2/3}$ [24], well outside experimental error. Anyway, it is clear that the gap is fully developed and $R_c^c$ << 1500 Å.

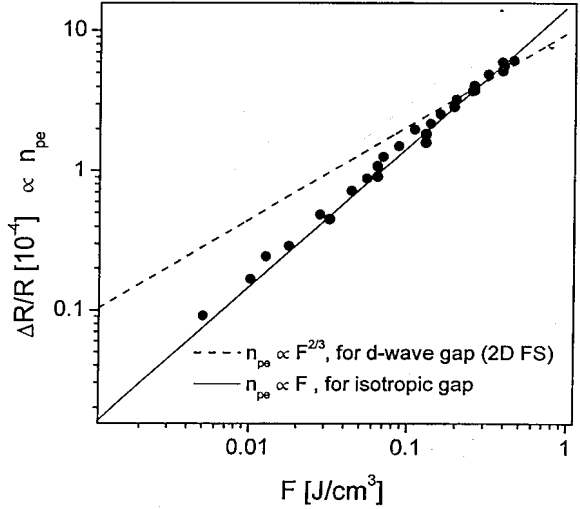

Fig.2: Ratio of the reflectivity change ΔR/R in optical pulsed experiments in YBCO as a function of the fluence F of the first pulse, the second second being the probing pulse. The data are from two different laboratories [25,26] plotted by J. Demsar. Included is also the expected fluence dependence F$^{2/3}$ **for** d- as well as F for s-symmetry.

No good photoemission data exist in YBCO; however, a large number of reliable experiments have been conducted in BSSCO. The outcome of practically all of them is a node in the superconducting gap along the Γ-M direction of the Brillouin zone [27,28]. This is a clear signature of d-symmetry. For the photoemitted carriers, the escape depth is one to at most two superconducting layers [27]. Thus the superfluid in the last plane near the vacuum is of d-symmetry. Assuming a similar optical picosecond reflection result as in YBCO, it can be speculated that $R_c^c$ is on the order of only a few unit cells. Practically all tunneling experiments along the c-axis also yield d-symmetry, to more than 90 percent [1, 29]. They probe only the last superconducting layer [2,28]. Therefore, from our estimates of $R_c^{ab}$ and $R_c^c$ above, we find that the bulk of the cuprate superconductors have s-character. *Morover this is the case for magnetic fields below $H_{c1}$, where no flux penetrates the sample* [30]. Fig. 2 is a clear example. An other is the transverse magnetic moment as a function of angle in the Meissner state. [31], discussed in Ref. 2.

A major objection against the proposed „d-outside and s-inside" model so far is the phase stiffness of the superfluid over a macroscopic sample [32] as probed by Josephson-junction and related experiments. Here one has to consider the inhomogeneity of the single crystals or epitaxial layers used. Because of this inhomogeneity there is, in our opinion, a percolative net of reduced superconductivity, comparable to an area near a surface, in which d-symmetry is predominant, ensuring the phase stiffness. To document this possibility, we summarize the recent effort by Schneider [33] for $Bi_2Sr_2CaCu_2O_{8+\delta}$. He used finite-size scaling theory to analyze some very accurate specific-heat data on the best-known single crystals from the University of Geneva. There the peak of $c_p$/ T near $T_c$ moves to *higher* temperatures upon application of a magnetic field parallel to the c-axis! His analysis (which is robust ) shows that the crystals are *inhomogeneous*, he calls them „granular". The „granularity" is 68 Å in size, in agreement with microwave data that he also analyses. Therefore in this case it is not so difficult to imagine a percolative „d-net" extending over the entire sample and maintaining the phase. YBCO and LSCO are certainly more homogenous. Thus the „d-net" is less dense, but apparently enough that the phase is coherent over macroscopic distances. In addition, at present it is an open question whether a d-component persists in the bulk which would help to assure the phase stiffness observed.

Before closing, it should be noted that the interacting boson model is compatible with supersymmetry [4]. The latter property is of relevance in the cuprates. for half a dozen years, the notion of a two-quasi-particle paradigm has imposed itself experimentally, one of the quasi-particles having fermionic, the other bosonic character [34,35]. Owing to the *fast dynamics* present in the ground state, transforming one kind into the other, supersymmetry might be present, which is compatible with the interacting boson model [4] as well as the proposed change of the superconducting wavefunction symmetry with increasing distance from the surface [2].

Acknowledgements: I am in debt to Jure Demsar for various discussions, and for letting me have Fig 2. before publication. I thank John D. Dow for sending me the preprint of Ref. 23, and acknowledge detailed correspondence with Richard Klemm on recent tunneling experiments along the c-axis for several cuprates, which could not be reviewed here because of the length limitation of this paper. I most appreciated the reading of the manuscript by Boris Kochelaev and his remark on the symmetry of the superconductors in the Meissner state.